\documentclass[lettersize,journal]{IEEEtran}
\usepackage{amsmath,amsfonts}
\usepackage{algorithmic}
\usepackage{algorithm}
\usepackage{array}
\usepackage[caption=false,font=normalsize,labelfont=sf,textfont=sf]{subfig}
\usepackage{textcomp}
\usepackage{stfloats}
\usepackage{url}
\usepackage{verbatim}
\usepackage{graphicx}
\usepackage{cite}
\usepackage{xcolor}
\begin{document}

\title{Axial Magnetic Quadrupole Mode of Dielectric Resonator for Omnidirectional Wireless Power Transfer}

\author{Esmaeel Zanganeh, Elizaveta Nenasheva, Polina Kapitanova
        % <-this % stops a space
\thanks{This work was supported by the Russian Science Foundation (Project No. 20-72-10090). PK acknowledges the Priority 2030 Federal Academic Leadership Program. EZ acknowledges Dr. Alena Shchelokova for sharing the ceramic hollow disk resonators. (Corresponding author: Esmaeel Zanganeh.)}

\thanks{Esmaeel Zanganeh and Polina Kapitanova are with School of Physics
and Engineering, ITMO University, 197101, Saint Petersburg, Russia. (e-mail: esmaeelzanganeh@gmail.com)

Elizaveta Nenasheva is with Ceramics Co., Ltd, 194223, Saint Petersburg,
Russia.}}

% The paper headers
\markboth
{{Zanganeh \MakeLowercase{\textit{et al.}}: Axial Magnetic Quadrupole for Omnidirectional WPT}}
\}

\maketitle

\begin{abstract}
 To achieve omnidirectional wireless power transfer with high efficiency, a high Q-factor transmitter generating homogeneous magnetic field is crucial. Traditionally, orthogonal coils of different shapes are used to realize transmitters. In this paper, we develop an omnidirectional magnetic resonant wireless power transfer system based on a dielectric disk resonator with colossal permittivity and low loss operating at axial magnetic quadrupole mode. The constant power transfer efficiency of $88\%$ at the frequency of 157~MHz over the transfer distance of $3~cm$ for all angular positions of a receiver is experimentally demonstrated. The possibility of multi-receivers charging is also studied demonstrating a total efficiency of $90\%$ regardless of angular position between two receivers with respect to the transmitting disk resonator.  The 
 %advantage of the proposed design is the
 minimized exposure of biological tissues to the electric and magnetic fields as well as a low specific absorption rate is observed that makes the WPT system safer for charging with higher input power.
\end{abstract}


\section{Introduction}
Wireless power transfer (WPT) technologies are used to charge batteries of different electronic devices \cite{Song2021,mahesh2021inductive, 8357386, Song2017, jiang2017overview}. Among different techniques, the magnetic resonant WPT has attracted great attention for its potential in safe mid-range charging \cite{kurs2007wireless,dionigi2022magnetic,niu2019state,mou2019survey,Song2020multi}. However, the  power transfer efficiency (PTE) of magnetic resonant WPT systems based on metal coils still suffer from high ohmic and radiation losses and their angular instability \cite{7497600,shamonina2021wireless,sasatani2021room}. To reduce the ohmic loss and enhance the PTE, the use of dielectric resonators instead of metallic coils have been recently proposed \cite{song2016wireless2,song2016wireless,PhysRevApplied.15.044024}. To further enhance the PTE of WPT systems, a dielectric metasurface supporting a quasi magnetic bound state in the continuum was introduced \cite{PhysRevApplied.15.044024}. Another useful feature of the dielectric resonators is the ability to engineer the response by combination of different modes \cite{shamkhi2019transverse,miroshnichenko2015nonradiating,babicheva2017resonant}. Thus, nonradiating sources based on anapole states of dielectric disk resonators can be obtained \cite{zanganeh2021anapole,kapitanova2020seeing,basharin2023selective,zanganeh2020electromagnetic} bringing a benefit for PTE enhancement of WPT systems due to suppression of radiation losses \cite{zanganeh2021nonradiating}.

To expand charging regions for more convenient utilization, modern WPT systems are required to be insensitive to positional misalignment \cite{kim2016free,Zanganeh2023Extreme}. Therefore, for WPT to arbitrary spatial position devices, omnidirectional systems have been intensively investigated \cite{feng2021load,en11082159,8683985,8676220,lin2016mathematic,ha2017analytical,han20193d,lu2020design,8989790,ha2022cylindrical,9723519,9374772,9492820}. Most of the proposed omnidirectional WPT systems are based on orthogonal transmitting (Tx) coils of different shapes \cite{8683985,feng2021load,8676220,en11082159}. Such structures effectively eliminate the cross-coupling effect between multiple 
%transmitters
coils and generate uniform magnetic field. However, these structures need several power sources to feed each Tx coil and dynamically control the phases and amplitudes of coil currents. Recently, Tx coils fed by a single power source generating uniform magnetic fields were implemented \cite{ha2017analytical,lu2020design}. One of the drawbacks of these systems is the existence of the blind zones with very small electromagnetic coupling to a reciever (Rx) and as a result, low PTE.  Another limitation is the safety of omnidirectional WPT systems to the biological tissues. As soon as the shielding techniques could not be applied, the level of electric and magnetic fields created by omnidirectional WPT systems must satisfy the limitations regulated by the standard \cite{ieee2019ieee}. Thus, the allowable power transferred by an omnidirectional WPT system could be limited.

In this letter, we propose an omnidirectional WPT system based on a dielectric hollow disk resonator operating at axial magnetic quadrupole mode. We demonstrate that at this mode, the resonator produces a homogeneous magnetic field in the transverse plane which can be potentially used to enable equivalent power transmission to Rxs at any angular position around the Tx. We theoretically and experimentally study the omnidirectional WPT system with single and two Rxs in MHz frequency band.  We demonstrate that the advantage of the dielectric resonator with respect to metallic coils is the confinement of the electric field. As a result, the exposure of electric and magnetic fields to biological tissues is minimized. This results in a very low specific absorption rate (SAR), making it a safer option for omnidirectional wireless charging.
\begin{figure}[t]
\centering
\includegraphics[width=8.2cm]{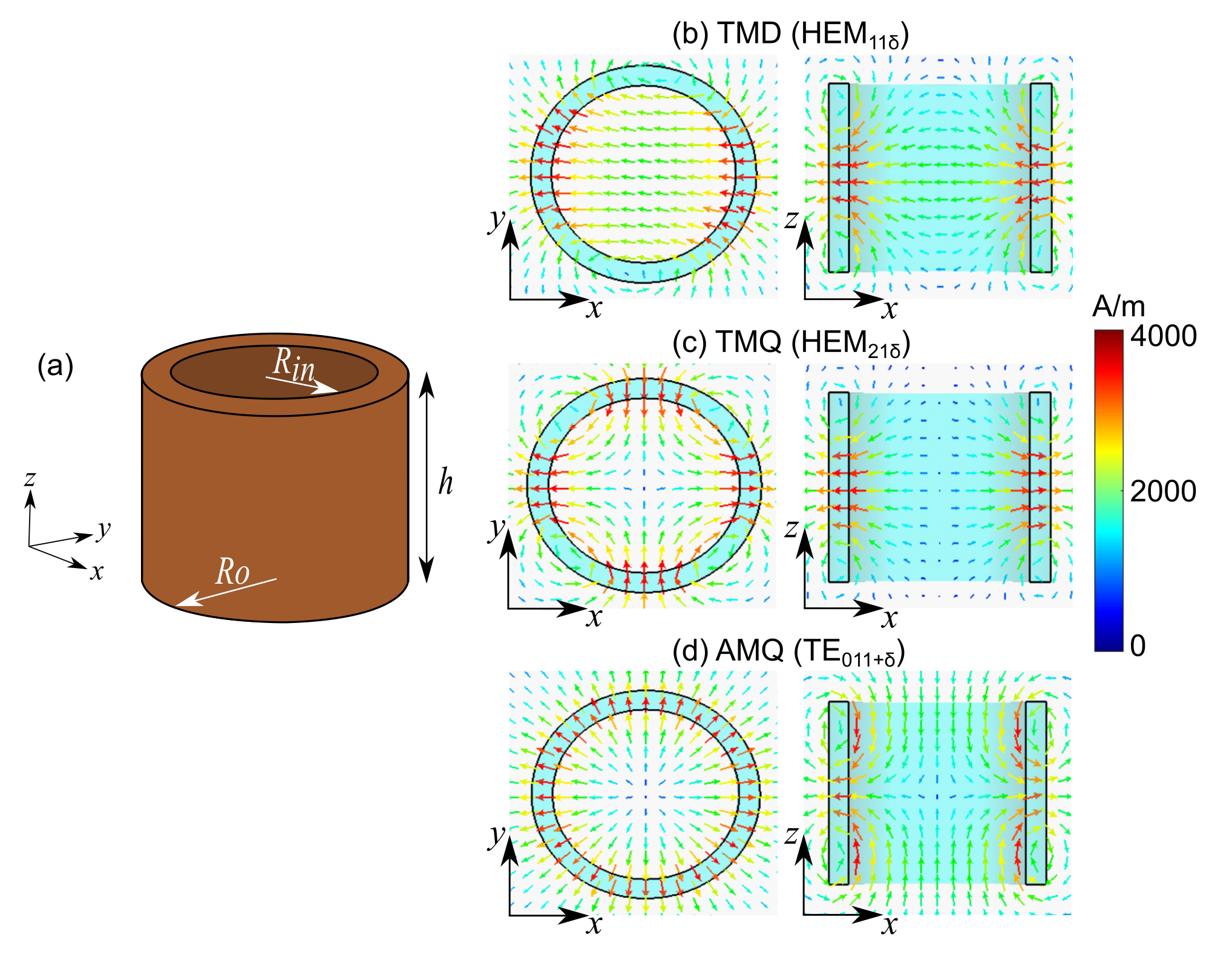}
\caption{(a) Schematic view of the dielectric hollow disk resonator with outer radius $R_{o}=62.2~mm$, inner radius $R_{in}=50.75~mm$, height $h=100~mm$, permittivity $\varepsilon=1000$ and loss tangent $tan\delta = 4\times10^{-4}$. Simulated magnetic field distributions of (b) the TMD (b), TMQ (c), and AMQ (d) resonator modes in the transverse and axial planes.}
\label{fig1}
\end{figure}
\begin{figure}[h]
\centering
\includegraphics[width=8.2cm]{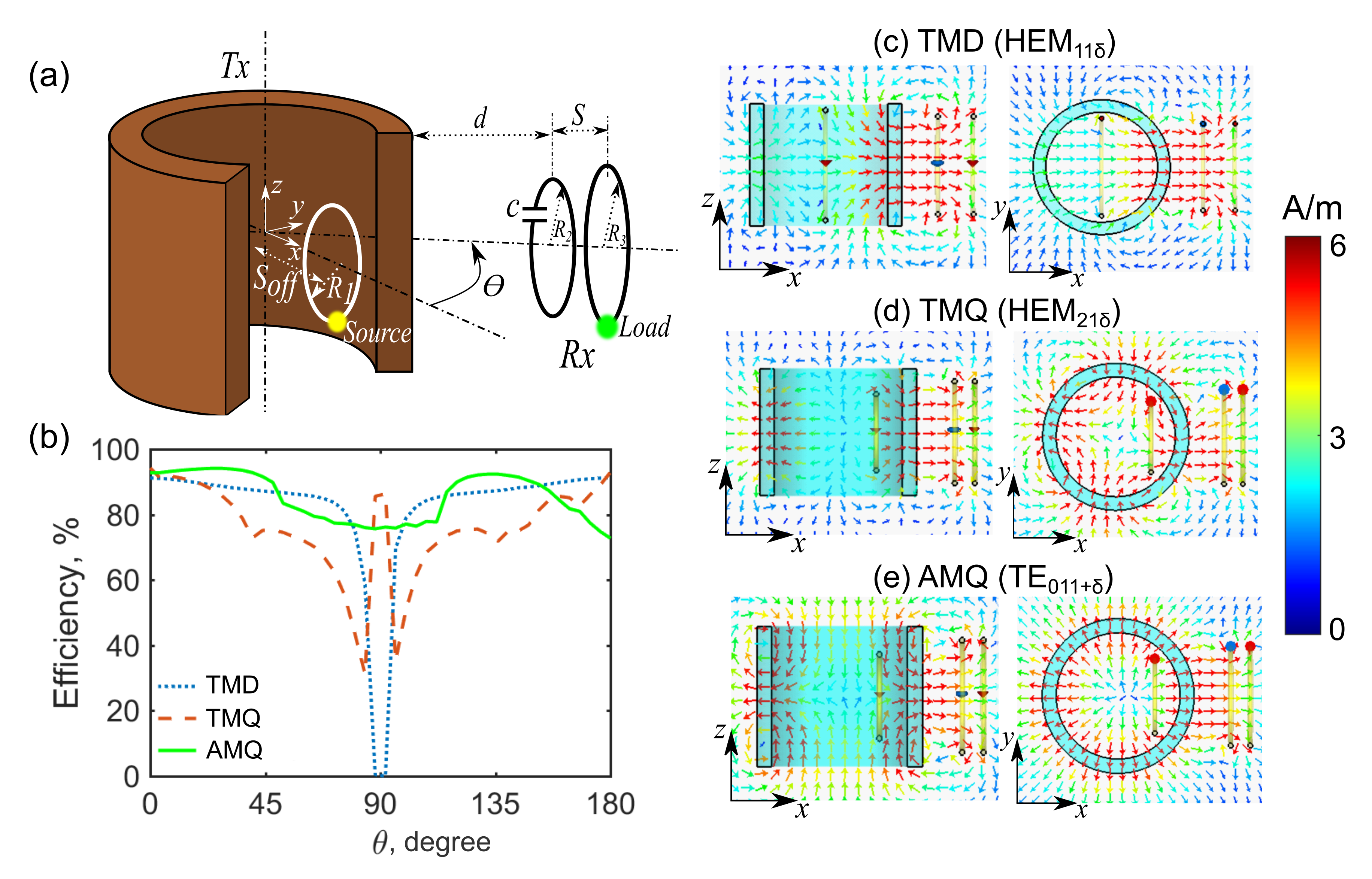}
\caption{(a) Schematic view of a WPT system based on the dielectric hollow disk resonator excited by a simple copper loop. The radius of the Rx resonator and load loop is set as $R_2 = R_3 = 40~mm$ for systems operating at all three modes under study. Transfer distance is $d=3~cm$. The wire radius is $t=2~mm$.(b) PTE of the WPT systems operating at the TMD, TMQ, and AMQ modes extracted from the simulated S-parameters by Eq(1). Simulated magnetic field distributions of the WPT systems operating at the TMD (c) TMQ (d), and AMQ (e) modes at the transverse and axial planes.}
\label{fig2}
\end{figure}
\section{Eigenmode analysis} 
A schematic view of the dielectric hollow disk resonator is shown in Fig.~\Ref{fig1}(a). We start with the eigenmode analysis of the dielectric resonator in CST Microwave Studio 2022, which reveals three modes at the frequencies of 138, 152.5, and 150.2~MHz, respectively. Simulated magnetic field distributions of all modes in both transverse ($\it{x}$-$\it{y}$) and axial ($\it{x}$-$\it{z}$) planes are illustrated in Fig.~\Ref{fig1}(b-d). For the first mode at the frequency of 138 MHz, the magnetic field oscillates along the $\it{x}$-axis (see Fig.~\Ref{fig1}(b)). From the point of view of traditional dielectric resonators antenna design, this mode is known as HEM$_{11\delta}$ \cite{mongia1994dielectric}. For the second  mode at the frequency of 152.5~MHz, the magnetic field is oscillating inside of the resonator providing four maximums in the transverse plane as shown in Fig.~\Ref{fig1}(c). This mode is classified as HEM$_{21\delta}$ mode. At the frequency of 150.2~MHz, the homogeneous radial magnetic field distribution in the transverse plane together by oscillating fields in z-direction is obtained. This mode is classified as TE$_{011+\delta}$ mode. However, using Mie scattering theory \cite{bohren2008absorption}, the obtained HEM$_{11\delta}$, HEM$_{21\delta}$, and TE$_{011+\delta}$ modes can be classified as transverse magnetic dipole (TMD), transverse magnetic quadrupole (TMQ), and axial magnetic quadrupole (AMQ), respectively.

To design a highly efficient omnidirectional WPT system, one needs to provide a high Q-factor and uniform magnetic field distribution over the angle. The numerically estimated Q-factor is found as high as 2510 for all the modes under consideration. However, the magnetic field distributions of TMD and TMQ modes have blind zones which may lead to weak coupling with Rx. The AMQ mode provides a uniform magnetic field distribution in the transverse plane. Being properly excited, this mode can be used to overcome the blind zone limitation and implement the Tx for an omnidirectional WPT system enabling power transmission to Rxs at any angular position around the Tx.
\begin{figure}[b]
\centering
\includegraphics[width=8.2cm]{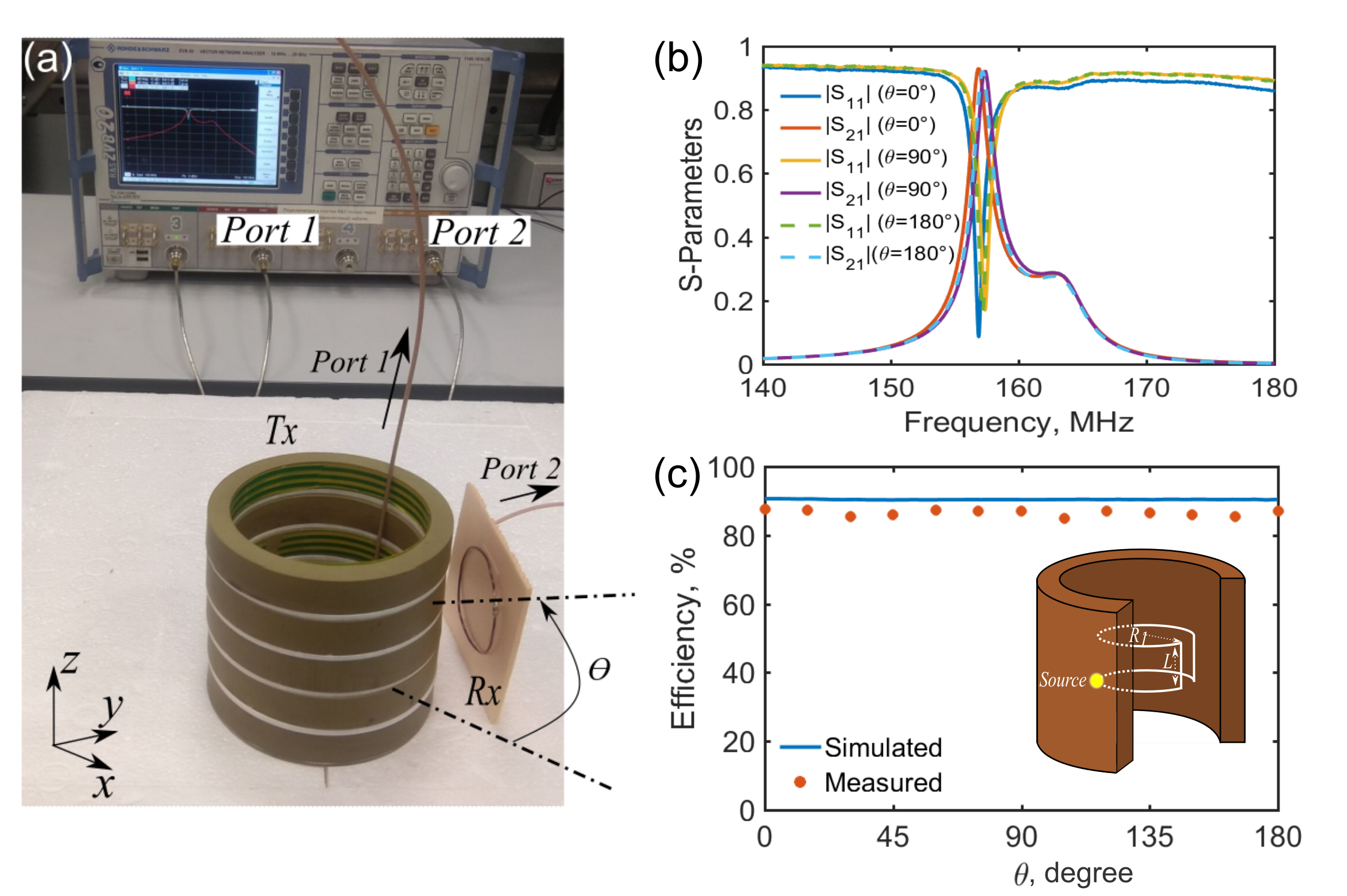}
\caption{(a) Photo of the omnidirectional WPT system experimental setup with one $Rx$. (b) Measured S-parameters of the system for different angular positions of the $Rx$. (c) Measured and simulated PTE of the omnidirectional WPT system  at the frequency of 157~MHz as a function of the $Rx$ angular position $\theta$. The inset shows the design of the $Tx$  with symmetrical excitation shielded loop centered on the hollow disk with a radius of $R_1=45~mm$ and height $L = 25~mm$ and a $2~mm$ gap between its two vertical sections.}
\label{fig3}
\end{figure}

\section{Numerical study} 
The PTE of WPT systems based on the dielectric hollow disk resonator operating at the TMD, TMQ, and AMQ modes is numerically studied. A schematic view of the WPT system created in the CST is shown in Fig.~\Ref{fig2}(a). The dielectric resonator excited by a copper source loop with radius $R_1$ is used as the Tx. The Rx resonator is placed at a transfer distance $d$ with respect to the Tx. The Rx resonator is a simple copper loop with a radius of $R_2$ terminated by a capacitor $C$ to provide a resonance at the same frequency as the Tx. A load loop made of a copper wire with a radius of $R_3$ is placed at a distance of $S$ from the Rx resonator. A $50~\Omega$ port is inserted in the slits on both source and load loops. The reflection 
%($|S_{11}|$) 
and transmission 
%($|S_{21}|$) 
coefficients for the different angular positions of the Rx are numerically obtained in 
%frequency-domain solver of 
CST Microwave Studio and used to calculate the PTE as follows:
\begin{equation}
	\eta =\frac{\mid S_{21} \mid^2}{1-\mid S_{11} \mid^2}\times100~\%.
\end{equation}
The geometrical parameters and capacitance value obtained during the numerical optimization of the WPT systems at different modes are listed in Table.~\ref{tab:table1}. The PTEs calculated by Eq (1) as a function of Rx angular position are shown in Fig.~\Ref{fig2}(b). At TMD mode the PTE is zero for the angles around $90^o$ and the Rx cannot be charged there. The magnetic field of the system working at TMD is oscillating in $\it{x}$ direction and is minimum around $\theta=90^0$. For the system working at TMQ mode, there are two blind zones around $80^o$ and $100^o$ angles, where the PTE is around $40\%$. The magnetic field is oscillating along both $\it{x}$ and $\it{y}$ axes (Fig.~\Ref{fig2}(d)). In the system working at AMQ mode, the PTE is over $80\%$ for all angles. The magnetic field provides almost uniform distribution in the transverse plane Fig.~\Ref{fig2}(e). However, the PTE is not still constant due to the asymmetric excitation of the Tx. 
\begin{table}[t]
%\centering
\caption{Geometrical parameters, capacitor value, and resonant frequency of WPT systems operating at TMD (HEM$_{11\delta}$), TMQ (HEM$_{21\delta}$), and AMQ (TE$_{011+\delta}$) modes.}
\begin{tabular}{lccccc}
\hline\hline
 \textbf{Mode} & $R_1$(mm)  &  $S_{off}$ (mm) & $C$ (pf) & $S$ (mm) & f (MHz)\\
  %\midrule
  \hline \textbf{TMD}  &45 & 0 & 7.4 & 15& 138\\
 \hline \textbf{TMQ}  & 30 & 30& 6.45 & 12 & 153 \\
 \hline \textbf{AMQ}  &30 &30& 6.7 & 12 & 150 \\
 \hline\hline
\end{tabular}
 \label{tab:table1}
\end{table}
To achieve the constant PTE of the WPT system working at the AMQ mode, we come up with a symmetric source loop design, as shown in the inset of Fig.~\Ref{fig3}(c). The PTE of the WPT system working at the AMQ mode with symmetrical excitation is calculated and shown in Fig.~\Ref{fig3}(c). The maximal PTE of $90\%$ for all angular positions of the single Rx is achieved numerically.

\section{Experimental study} 
 Next, we experimentally study the performance of the WPT system operating at the AMQ mode. The fabricated prototype of the WPT system is shown in Fig.~\Ref{fig3}(a). It comprises of a dielectric hollow disk resonator excited by a  symmetric source loop as the Tx, and an Rx resonator coupled to a load loop. Due to some technical limitations, we could not fabricate the disk with the parameters used in simulations. Thus, we stack it of five hollow disks with the following dimensions: the inner radius of $R_{in} = 50.75~mm$, the outer radius of $R_o = 62.2~mm$, and the height of $h_d = 20~mm$. The hollow disks are made from BaSrTiO3, including Mg-containing compositions and have the permittivity of $\varepsilon\ = 1000$ and $tan \delta = 4\times10^{-4}$ (at 1~MHz) \cite{nenasheva2010low}. Since the ceramic disks are fragile, several thin hollow spacers made of plexiglas with relative dielectric permittivity around $3.5$ and electrical conductivity of $0.02~Sm^{-1}$ and height of $3~mm$ are used between them.  The symmetric source loop is a  shielded loop made of a coaxial cable \cite{carobbi2004analysis}. 
The Rx resonator is made of copper wire with radius $R_2 = 34~mm$ and wire diameter $t =1~mm$. To provide the resonance at the frequency of 157 MHz, the Rx is matched by $C=6.8~pf$ capacitor. The load loop radius is $R_3=30~mm$. The end of the source and load loops are connected to the $50~\Omega$ ports of a Vector Network Analyzer (VNA) by coaxial cables. 

The S-parameters of the WPT system as a function of the Rx angular position with the step of $15^o$ are measured and several of them are shown in Fig.~\Ref{fig3}(b). Then, the PTE is calculated using Eq.(1) and compared to the simulated PTE in Fig.~\Ref{fig3}(c). An equivalent PTE above $88\%$ for all Rx angular positions is achieved.

\begin{figure}[t]
\centering
\includegraphics[width=8.2cm]{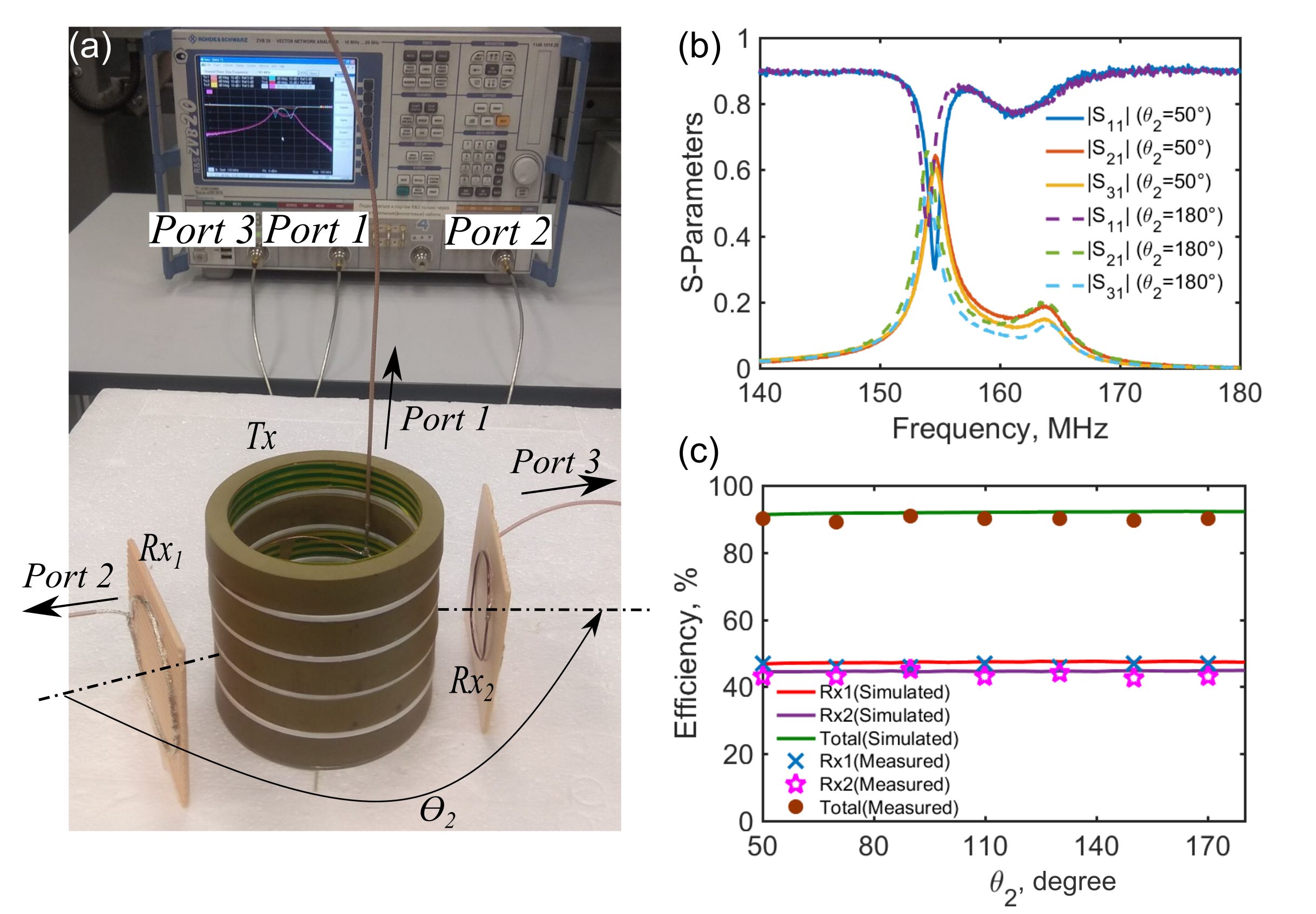}
\caption{(a) Photo of the omnidirectional WPT system experimental setup with two $Rxs$ separated by angle $\theta_{2}$. (b) Measured S-parameters of the omnidirectional WPT system for $\theta_{2}=50^0$ and $\theta_{2}=180^0$. (c) Measured and simulated PTEs of $Rx_1$, $Rx_2$ and total PTE of the omnidirectional WPT system  at the frequency of 154~MHz as a function of $\theta_2$. Note that the angles between $0^o$ to $50^o$ are not feasible due to the receivers’ size restrictions.}
\label{fig4}
\end{figure}

\begin{table*}
\centering
\caption{Comparison of omnidirectional and directional magnetic resonant WPT systems based on metallic and dielectric resonators.}
\begin{tabular}{lccccccc}
\hline\hline
 \textbf{Ref.}  & Frequency  &  Tx resonator type  & PTE($\%$) & &       Blind zone& Distance/ $\lambda$ &  Tx size/ $\lambda$\\
  %\midrule
\hline \textbf{\cite{kurs2007wireless}}  & 10~MHz &   Metal coil   &     94&   &   Yes&   0.025 &   0.02\\
\hline \textbf{\cite{song2016wireless2}}&  232~MHz&  Dielectric disk   &       90&    &   Yes &    0.03 &    0.0649 \\
\hline \textbf{\cite{song2016wireless}} &2.4~GHz&  Dielectric sphere  &        80&        &   Yes & 0.05 &   0.8 \\
\hline \textbf{\cite{zanganeh2021nonradiating}}  & 408~MHz  & Dielectric disk  & 92&  & Yes& 0.055 &  0.1142 \\

\hline \textbf{\cite{lin2016mathematic}} & 535~KHz  &Orthogonal coils &78 & & No & 0 &  $5.35\times10^{-4}$ \\
\hline \textbf{\cite{ha2017analytical}} &  13.56~MHz  &Cubic Metal coil  &60&    & No &0.0136 &  0.009\\
\hline \textbf{\cite{han20193d}}  & 20~KHz  &Orthogonal coils & 11.5 &  &No & $4\times10^{-6}$  &  $2\times10^{-5}$\\
\hline  \textbf{\cite{lu2020design}}   &  13.56$\And$27.12~MHz  &  Metal coil  &  61.6$\And$65.4 &  &   No&  0.004 &  0.0072\\
\hline \textbf{\cite{ha2022cylindrical}}  &  6.78~MHz  & Metal coil & 74.2 (DC-DC) &  & No&$4.5\times10^{-4}$ &  0.0034\\
\hline \textbf{This work}  & 157~MHz  &Dielectric disk & 88&     & No& 0.016 &   0.065 \\
\hline\hline
\end{tabular}
 \label{tab:table2}
\end{table*}

\section{Multi-receiver omnidirectional WPT} 
 We also experimentally study the system with two Rxs. An identical second Rx ($Rx_2$) is added to the system with the angular position of $\theta_2$ with respect to the $Rx_1$ as shown in Fig.~\Ref{fig4}(a). The S-parameters of the system as a function of the $\theta_2$ with the step of $20^o$ are measured and several of them are depicted in Fig.~\Ref{fig4}(b). One can see that not only the operational frequency is stable for all angular positions, but also the transmission and reflection coefficient magnitudes are at the same level. The PTEs of $Rx_1$ and $Rx_2$ are calculated from the measured S-parameters using Eq (1). The total PTE is calculated as the sum of the first and second Rx PTEs.

The variation of the $Rx_1$, $Rx_2$, and total PTEs concerning the angle between the $Rx_1$ and $Rx_2$, ($\theta_2$), are compared to the simulated results in Fig.~\Ref{fig4}(c). One can see the total PTE remains almost $90\%$ regardless of the addition of the $Rx_2$, or its position. Also, the PTEs of $Rx_1$ and $Rx_2$ remain stable and almost equal for all angles. Therefore, it is clear that the WPT system operating at AMQ can be applied to omnidirectional WPT with multi-receivers.

 \section{Discussion} 
To understand the benefits of the proposed WPT system, we compare its characteristics with other implementations reported in the literature (Table.~\ref{tab:table2}). The proposed omnidirectional WPT system operates at the frequency of 157~MHz and achieves $88\%$ of the PTE to a single Rx over $3~cm$ transfer distance. This PTE is comparable to the systems reported in \cite{kurs2007wireless,song2016wireless2,zanganeh2021nonradiating,song2016wireless} that do not offer omnidirectionality.  Compared to the omnidirectional WPT systems reported in \cite{lin2016mathematic,ha2017analytical,han20193d,lu2020design,ha2022cylindrical}, the proposed WPT system provides the highest PTE over a larger distance without any blind zone. For instance, our proposed system outperforms the metal coil-based system in Ref.~\cite{lu2020design} in both efficiency ($88\%$ vs. $65.4\%$) and operation distance normalized to wavelength ($0.016$ vs. $0.004$). This demonstrates the superior performance of our proposed WPT system.

\begin{figure}[h]
\centering
\includegraphics[width=8cm]{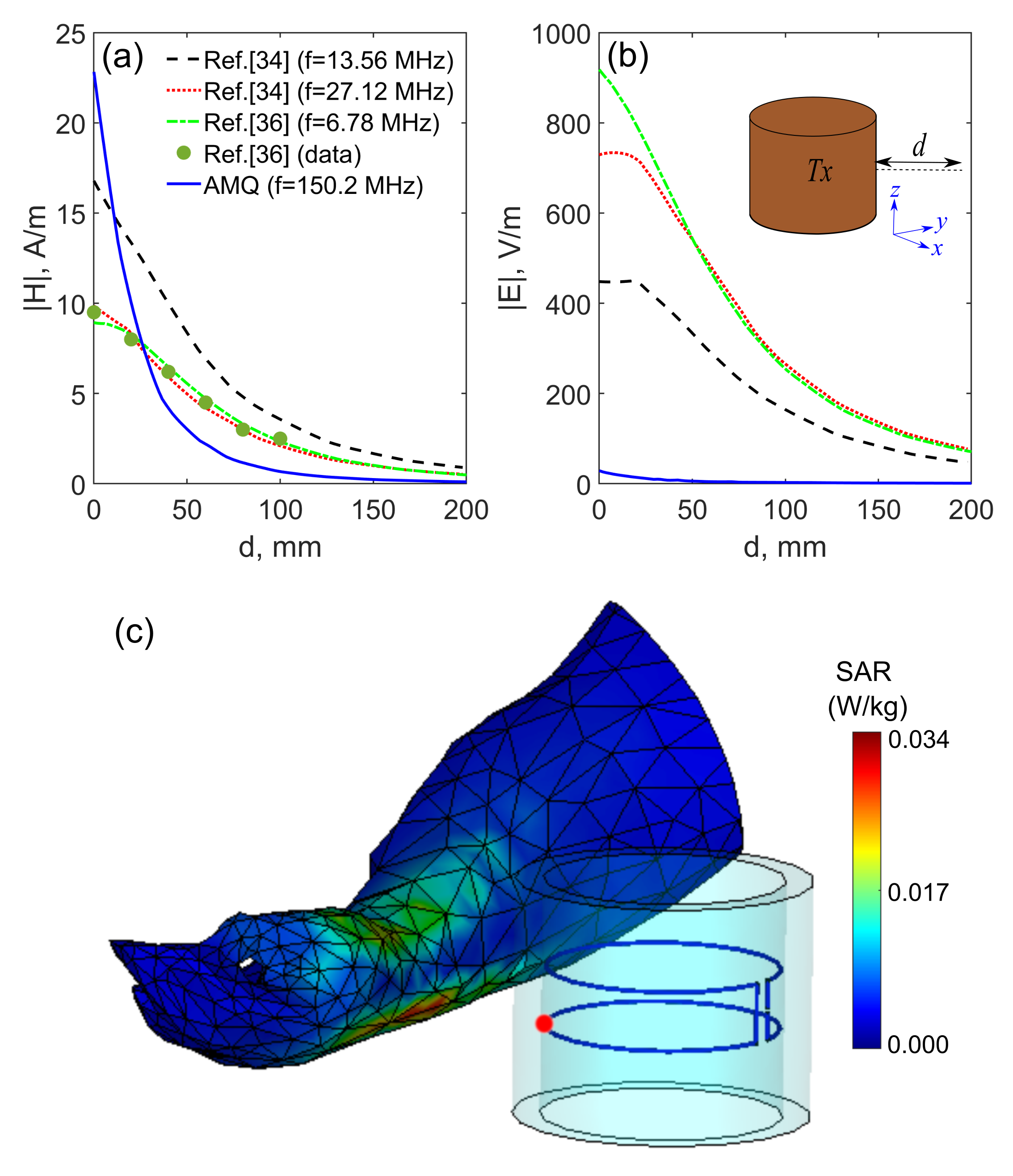}
\caption{ Simulated (a) magnetic and (b) electric fields of the proposed omnidirectional WPT Tx as a function of the distance $d$ compared to the Tx of the omnidirectional WPT systems based on metallic coils reported in Refs~\cite{lu2020design} (square coil) and \cite{ha2022cylindrical}. 
To verify the correctness of our simulations with the results presented in Ref.~\cite{lu2020design} we made sure that the reflection coefficient coincides with Fig. 12(a) in Ref.~\cite{lu2020design}. Magnetic field data presented in Fig. 2 of Ref.~\cite{ha2022cylindrical} is added to panel (a), which coincides with the recalculated magnetic field of the Tx. 
(c) Simulated SAR of the omnidirectional WPT Tx in an arm located at a distance of $4~cm$.}
\label{fig5}
\end{figure}

 To ensure the safety, it is crucial to examine the field exposure to biological tissues and SAR~\cite{ieee2019ieee}. The electric and magnetic fields of the proposed WPT system's Tx with symmetric excitation are compared with fields of the metal Txs of omnidirectional WPT systems reported in Ref.~\cite{lu2020design} and Ref.~\cite{ha2022cylindrical}. The central cross sections of the electric and magnetic field distributions simulated in CST Microwave Studio are compared in Fig.~\ref{fig5}(a) and (b), respectively.
The magnetic field of the proposed Tx based on dielectric disk resonator has the highest amplitude at short distances $d$ in comparison to the magnetic fields provided by the designs reported in Ref.~\cite{lu2020design} and Ref.~\cite{ha2022cylindrical}. Its magnitude decays faster as $d$ increases (Fig.~\ref{fig5}(a)), indicating strong accumulation of the magnetic field in proximity of the dielectric resonator. The electric field of the proposed Tx (see Fig.~\ref{fig5}(b)) is negligible compared to electric fields of the metal Txs reported in Ref.~\cite{lu2020design} and Ref.~\cite{ha2022cylindrical}. Therefore, the proposed omnidirectional WPT system based on the dielectric resonator offers the strong confinement of the electric field providing less exposure of the electric field~\cite{ieee2019ieee} to surrounding biological tissues compared to WPT systems based on metal Txs.

 To perform SAR analysis, we employ CST Microwave Studio and a computer-aided-design model of the front part of a human arm as depicted in Fig.~\ref{fig5}(c). The model consists of the main biological tissues of the arm characterized by their corresponding electromagnetic properties. The distance between the Tx and the arm is 4~cm. For the input power of 0.5~W, the maximal calculated SAR is 0.034~W/kg, averaged over 1~g of tissue (see Fig.~\ref{fig5}(c)). There are no nonlinear effects in the WPT system, and the maximal SAR for different input powers can be obtained by scaling up these results. According to the IEEE safety regulation \cite{ieee2019ieee}, which specifies the maximum SAR value for limbs and pinnae as 4~W/kg, a maximal input power of 117~W is allowed.

\section{Conclusion} 
We proposed the omnidirectional WPT system based on a dielectric hollow disk resonator operating at AMQ  mode.
A uniform radial magnetic field in the transverse plane of the Tx resonator is generated helping to avoid the blind zone. A symmetric excitation loop to achieve a constant PTE over all angular positions of the Rx is proposed. With respect to the experimental data, the omnidirectional WPT system provides $88\%$ of PTE over all angles at the transfer distance of $3~cm$ to a single Rx. The possibility of charging multi-receivers is also experimentally verified. The results showed a stable and high efficiency of $90\%$ regardless of the angle between two Rxs.  Examination of the safety issues of the proposed WPT system revealed minimal exposure of the electromagnetic fields to biological tissues, resulting in a very low SAR. These findings establish the proposed WPT system as a safer option for omnidirectional wireless charging.
%\section{Acknowledgements.} 
\bibliography{Main}% Produces the bibliography via BibTeX.
\bibliographystyle{IEEEtran}
\vfill
\end{document}